\documentclass[conference]{IEEEtran}
\usepackage{cite}
\usepackage{amsmath,amssymb,amsfonts}
\usepackage{algorithmic}
\usepackage{graphicx}
\usepackage{textcomp}
\usepackage{xcolor}
\usepackage{multirow, booktabs}
\usepackage{listings,multicol}
\usepackage{xfrac}
\usepackage[utf8]{inputenc}
\usepackage[T1]{fontenc}
\usepackage{fixltx2e}
\usepackage{csquotes}
\usepackage{amsmath,diagbox,tabu}
\usepackage{hyperref}
\usepackage{comment}

\begin{document}

\title{The Challenges of Investigating Cryptocurrencies and Blockchain Related Crime}

\author{\IEEEauthorblockN{Simon Dyson}
\IEEEauthorblockA{The Cyber Academy,\\
Edinburgh Napier University, Edinburgh. UK\\
Email: simon.dyson@protonmail.com}
\and
\IEEEauthorblockN{William J Buchanan, Liam Bell}
\IEEEauthorblockA{Blockpass ID Lab,\\
Edinburgh Napier University, Edinburgh. UK\\
Email: w.buchanan@Napier.ac.uk}
}

\maketitle

\begin{abstract}
We increasingly live in a world where there is a balance between the rights to privacy and the requirements for consent, and the rights of society to protect itself. Within this world, there is an ever-increasing requirement to protect the identities involved within financial transactions, but this makes things increasingly difficult for law enforcement agencies, especially in terms of financial fraud and money laundering. This paper reviews the state-of-the-art in terms of the methods of privacy that are being used within cryptocurrency transactions, and in the challenges that law enforcement face.
\end{abstract}

\IEEEpeerreviewmaketitle

\section{Introduction}
Mergenovna et al \cite{sat2016investigation} defined the increasing challenges of investigating money laundering and financing terrorism, as cryptocurrencies are increasingly used to hide the tracks of these crimes. Some criminals have even defended their actions against criminal activity by defining that the transactions were not of a financial nature \cite{mcleod2017bitcoin}. While many current cryptocurrencies provided pseudo-anonymous identifiers, several are now developing anonymisation layers which hide both the sender the recipient and the cost of a transaction. This anonymisation will make life increasing difficult in detecting and investigating a range of crimes. 

As we have public blockchains, the strive for anonymisation in both the transaction and processing is a key element for protecting privacy and preserving consent. Monero is one currency which has taken a lead on this, and which uses ring signatures and stealth addresses. There is also a rise of anonymised processing, such as with zk-Snarks \cite{reitwiessner2016zksnarks}. These mechanisms now support the hiding of the core information of a transaction, but where it is possible to not double spend or spend more currency that has been allocated to a user.

\section{Background}
In a traditional finance infrastructure, Bob trusts his bank (Bank A) and Alice trusts her bank (Bank B). A transfer of funds involves Bob finding out the identifier of Alice's bank (such as their sort code) and for her account identifier. The transfer of funds then involves him informing his bank that he wants to transfer the funds to Alice (Figure \ref{fig01}). Bob's bank then checks the transaction, and if it is valid, his account will be debited by the defined amount. His bank will then forward the transaction to Bank B, and where Alice's bank will credit her account. In this way, both Bank A and Bank B have a ledger which can be checked for the transaction. This method works well in investigating crime, as each bank can report on Bob and Alice's transactions, and also if they see any unusual transactions.

The cyberpunks of the 1990s started to question the requirements for banks to provide the intermediate exchange, especially in the profits that banks made from the transactions \cite{maurer2013perhaps}. Their approach was to use a publicly available ledger - the blockchain - and then sign for transactions with public key encryption. Miners could then compete to create a consensus for the recent transactions and the winner would add a block onto the blockchain.

\begin{figure}
  \includegraphics[width=\linewidth]{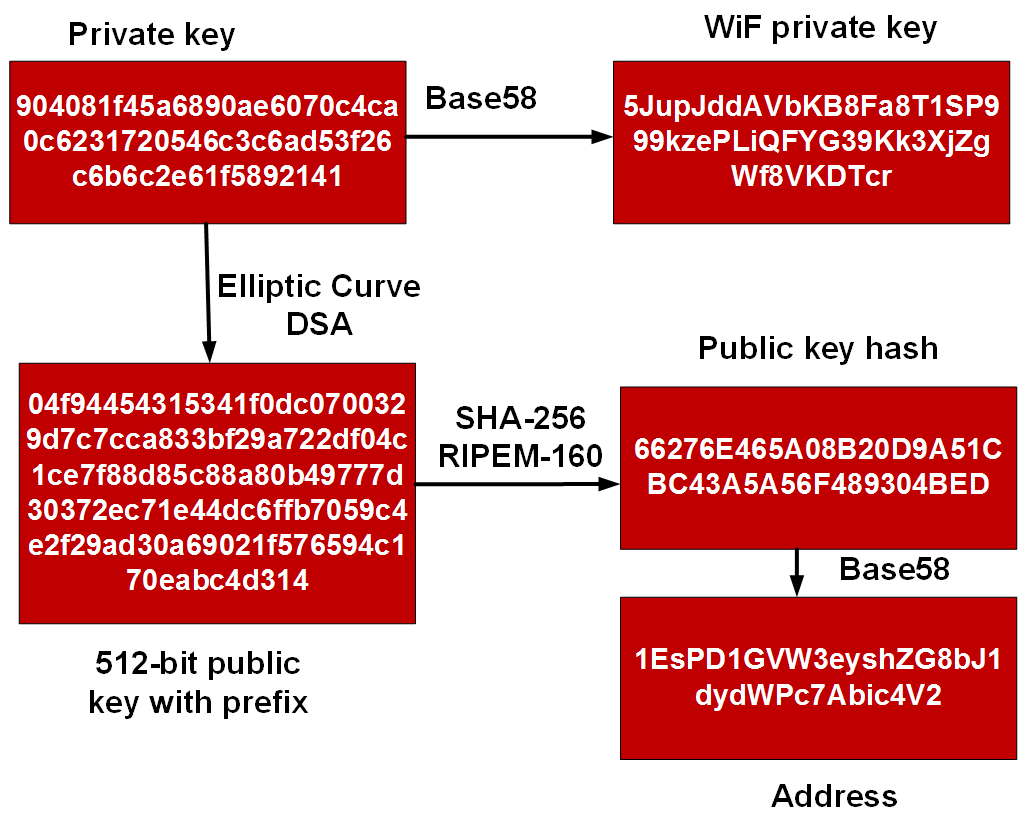}
  \caption{Bitcoin address creation}
  \label{bitcoin}
\end{figure}

\begin{figure}
  \includegraphics[width=\linewidth]{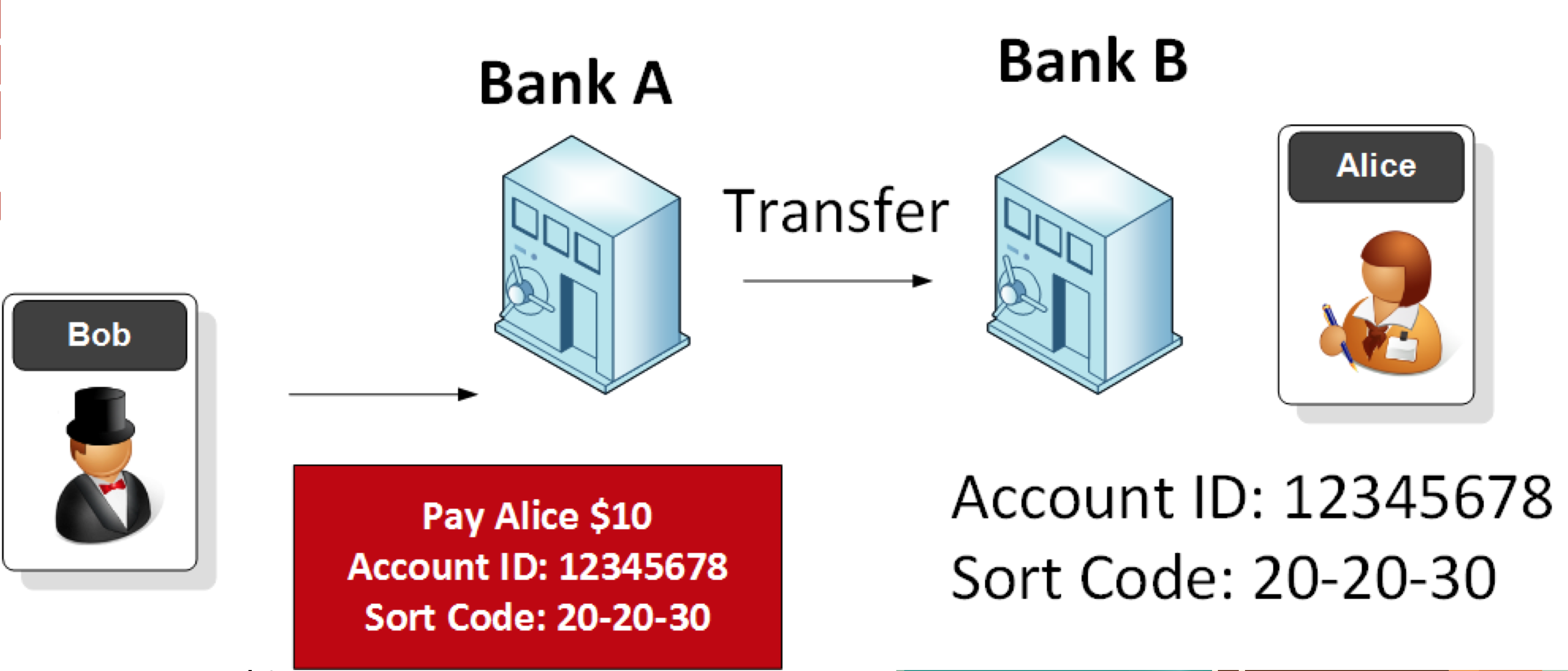}
  \caption{Traditional banking model}
  \label{fig01}
\end{figure}

\begin{figure}
  \includegraphics[width=\linewidth]{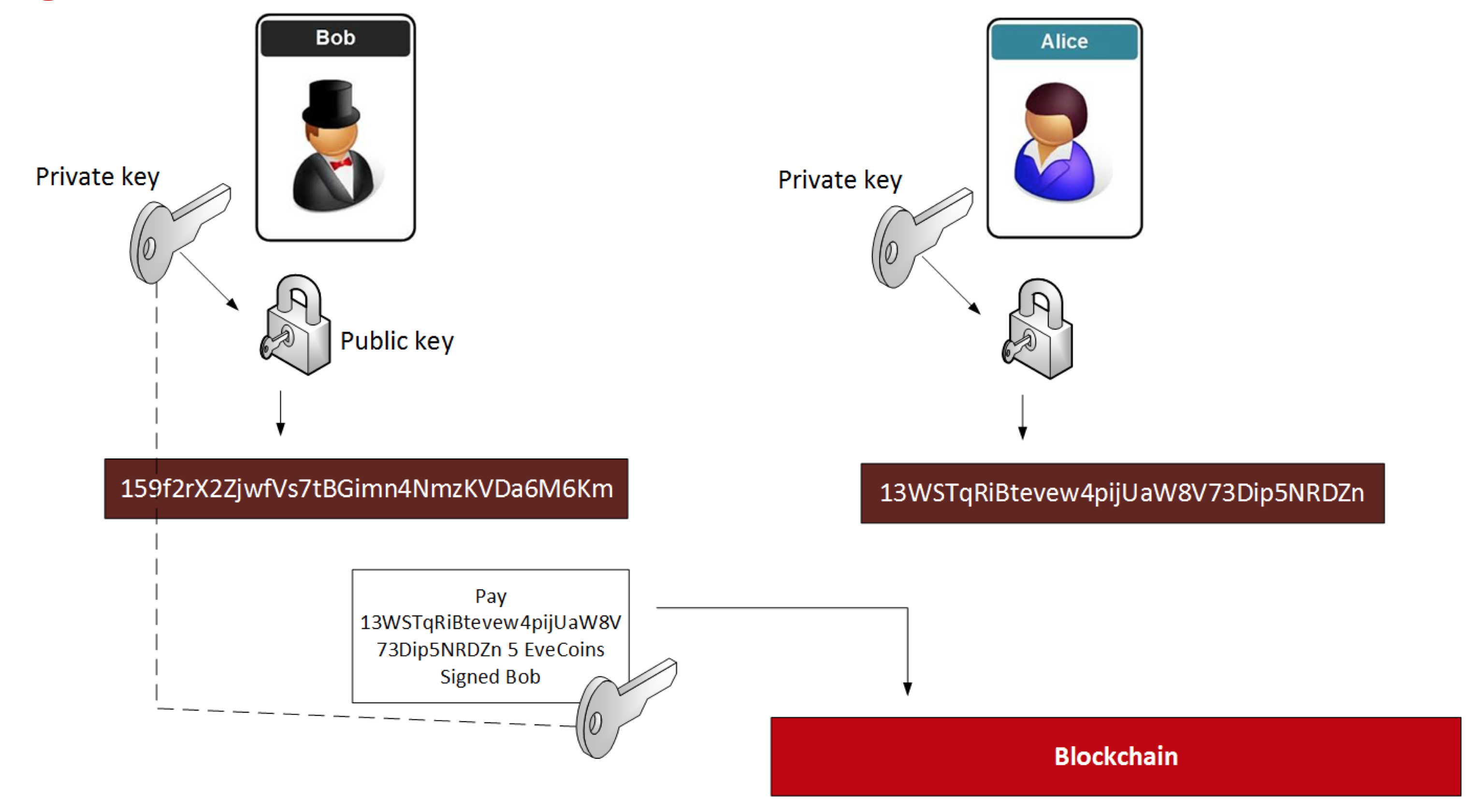}
  \caption{Basic blockchain process}
  \label{fig02}
\end{figure}

Within the Bitcoin infrastructure, Bob and Alice each generate a private key and then derive an associated public key. This public key is then used to create a public identification address for transactions (Figure \ref{bitcoin}). When Bob now wants to send Alice some funds, he determines her public address and then creates a transaction to send her a number of bitcoins (Figure \ref{fig02}). This is then signed with his private key and then picked up by miners who will gather together all the other recent transactions, and create a consensus for the transactions to be added to a new block on the blockchain. Before this can happen, the transaction needs to be checked to see if Bob has enough bitcoins in his account to pay Alice. This checking is the reason that the transactions need to be public, as the miners cannot process the transaction if Bob does not have enough funds to pay Alice. 

At the time of the creation of the bitcoin network, there were no feasible methods which could hide the fact that Bob was the payer and Alice was the payee. A pseudo-ID is then used to match Bob and Alice to a public address. While difficult to match the identifiers, law enforcement can at least trace is known addresses for their transactions. A worry with this model, though, is that the funds will never hit a bank account unless there is a cash out for funds into a fiat currency. This type of approach thus worries both tax raising authorities and law enforcement. For this reason, many governments around the world are now looking to regulate for cryptocurrencies, and thus provide an opportunity to audit their flows. A concern would be that it is possible to over regulate, and thus stifle innovation.

\section{Investigator challenges}

\subsection{Introduction} 

The new challenge for investigators is to understand the technical process that occurs during the transaction methods for each cryptocurrency algorithm. This understanding could range from the difference in the transaction system for Bitcoin-style systems to the state-based smart contract systems of Ethereum. The advancement in the methods used also increases the challenge for investigation, as Ethereum moves towards state channels, Plasma, Sharding, and upgrades to EWASM (Ethereum flavored WebAssembly) from the EVM (Ethereum Virtual Machine). There are also a number of challenges that investigators will need to admit defeat on and work with companies that have deeper visibility and resources to solve problems. Chainalysis \cite{chainalysis} and Elliptic \cite{Elliptic} are examples of market leaders in the field of cryptocurrency investigation and monitoring. They use addressing information that they are able to obtain that is not visible to an investigator using block explorers and OSINT (Open-source intelligence) techniques. This information is provided in products they provide that can point to services where further investigative leads can be explored. Law enforcement would have to build bespoke infrastructure to scrape and construct complex data structures to have visibility of this kind. 

The emergence of new crypto-economies is greater than those of currency and cash out. The ability to buy real estate or hire property for rentals shows how far this \emph{internet money} has come in a relatively short period of time. Purchasing of consumer goods is possible through direct purchase, intermediaries, gift schemes / Cards, debit cryptocurrency cards. Although as wildly reported the licence for Wavecrest was removed by Visa, disabling huge numbers of cryptocurrency debit card schemes that had allowed transfers of cryptocurrency to cards. There are however a number of similar alternatives still available worldwide and are likely to emerge with stronger KYC (Know Your Customer)/AML (Anti-money laundering) T\&Cs. As projects look to allow cryptocurrency to pay using applications with nearfield technology the future will likely see a disruptive innovation take hold \cite{Huang2018Chinese}. This already diverse use of assets details only a fraction of the usage that is possible now and certainly in future blockchain use cases. Localbitscoins also deals in peer to peer sharing that offers the exchange models of an online user to online user but also the ability to meeting in the street to exchange cash for cryptocurrency. This enables the ability for more users to become engaged and purchase but it is not without its flaws as those subject to fraud and deceit would attest.

\subsection{Centralised exchange AML/KYC region}

A cryptocurrency exchange generally can be considered a centralised exchange. This is controlled by a central operator a business. The exchange will hold the private keys and allows users to create accounts and pay in FIAT currency for cryptocurrency. Cryptocurrency exchanges range in the services that they offer to consumers. In basic terms purchasing from FIAT, US dollars to crypto (cryptocurrency) and the ability to exchange one cryptocurrency to another, such as swapping BTC Bitcoin to ETH Ethereum for a percentage fee. Many centralised services have more advanced services built into the platforms either as built-in additional options or as other products. These additional services mirror products that are associated with FIAT trading such as leveraging, margin lending, advanced stop loss and profit taking options. The exchanges provide advanced graphing and metrics for the currency pairings. 

Exchanges have changed considerably over the last few years in its adoption of KYC/AML (Know Your Customer / Anti-Money Laundering). It was possible to sign up for an account and cash in and out with very little or no details retained by exchanges other than an e-mail address. Exchanges, however, are now very large profitable businesses so it makes sense to apply due diligence and align to similar financial institutions. The space is unregulated and cryptocurrency has a shadow of illicit use from Silk road and its early use on the dark web. The introduction of self-regulation and due diligence is certainly a step in the right direction as the industry expands and large financial institutions are investing, interacting and in some cases, are now operating in the cryptocurrency financial industry. KYC now generally includes e-mail addresses, government issued photo id, home utility bills and often a photo holding the ID. Centralised exchanges are split into those based in KYC/AML friendly countries with high standards of AML legislation that will operate to high standards. Exchanges operating in these areas are aware that failure to adhere to expected standards will raise the spectre of compulsive regulation. Centralised exchanges expose the customers to risk if operated in a reckless manner such as BTC-E and MtGox. As these exchanges pull in vast amounts of customers as demonstrated by the relatively new exchange Binance, announcing higher profits than Germany's top bank making 200 million dollars in the same period.

\subsection{Centralised exchange non-regulated area}
 The non-regulated regions offer the same services discussed in the previous sections. The difference is that in regions where traditional money laundering occurs it is highly likely that these regions will be preferred for crypto laundering. If the region will not enforce or operate with law enforcement with international financial institutions then it is unlikely to push up the standards for this new financial adoption. These regions will also be influenced by government control, corruption, influence, destabilised governments or inefficient government regimes. In the coming years, it will be telling as investigations unravel across the globe to where the dirty cash falls out of the system. Will the traditional money laundering regions retain the higher rates of fraudulent activity.

\subsection{Decentralised exchange}
Centralised exchanges hold the private keys for the users so they are not secure and lack complete control for the user. In contrast decentralised exchanges offer a service that enables peer to peer trading without the need for an intermediary infrastructure. The loss of funds from centralised services has been observed in a number of attacks from direct theft from servers to social engineering, phishing, and brute force password attacks on the user accounts. There are a number of variations on decentralised exchanges where the infrastructure and appearance offers a peer to peer trade but would not be a truly decentralised exchange \cite{sat2016investigation}. There are multi chain peer to peer trading pairs that allow Bitcoin and Ethereum trading. True Ethereum / ERC20 token decentralised exchanges are hosted as Ethereum Dapps and complete the exchange using smart contacts. 

Atomic swaps are cross chain interactions that guarantee either a safe transaction or complete refund to both parties. This allows a smart contract exchange that posts a deposit and then an exchange where both parties cryptocurrency is contained in the system. This design is essential as there is no risk to funds from the collapse of the exchange, server breach or bad actors. 

A decentralised exchange can perform the functions of an exchange trading with the ability to create orders for buy and sell with advanced features. 

\subsection{Swaps, pairs and Shape-shifting}
There are a number of services that allow the exchange of tokens from one to another Shapeshift is one of the largest services of this type. The service has numerous methods of exchange from its own web service, partnerships with wallet providers and merchant services. The service doesn’t require KYC information identifier or an account it is a simple exchange. Criminal use of this service is documented in high profile cyber incidents such as WannaCry and the DAO hack \cite{Renaudin2018Truth}, \cite{Marks2018Crypto}.

There are a number of exchanges that operate these types of basic trading swap such as Changelly or Coinmotion. These services operate with low user friction and allow chain hopping, swapping out from Bitcoin to Ethereum or Monero for example. This chain hopping is often used to obfuscate and frustrate law enforcement activity. Shapeshift have however recently implemented a new membership token that includes the on-boarding of customers and trading limits for unregistered members. This appears to be more of a move towards a KYC/AML stance with incentives of better rates and features \cite{Leising2017Ether}.

\subsection{Mixers tumblers and fogging}

The use of mixers and tumbling services have become somewhat less popular than at the peak of the Silk Road market. These are still likely to remain as a constant as there is a still a requirement by some users to add additional layers of security and privacy. The effectiveness of mixers is questionable and certainly is costlier for those users submitting to these services. There is a number of tumblers and mixers using different methods to achieve obfuscation \cite{Voorhees2018Introducing}\cite{Bonneau2014Mixcoin:}\cite{Ruffing2014CoinShuffle:}. 

\subsection{Cryptocurrency Betting and gambling}

Money laundering has traditionally used betting and gambling in order to place money back into the system to move as legitimate money. This model is still used in the cryptocurrency and a large number of gambling sites and services. These include the distribution of illegally obtained funds to compromised gambling services such as a darknet world cup gambling ring in China \cite{Heilman2016TumbleBit:}.

\subsection{Cryptocurrency (ATM) Automated Teller Machine}
    
There are a number of crypto-currency ATM machine networks that allow for the debit and credit of cryptocurrency. These support Bitcoin and a number of alternative currencies such as Ethereum and Litecoin they are however still relatively small in number and use. 

\subsection{Decentralised market places}

There are new emerging decentralised market places that aim to replace “E-Bay” and Amazon based on decentralised networks and operating without centralised services. These new marketplaces offer customers safe peer to peer trading of goods without dissolving their rights to privacy and targeted advertising. The purchase of goods through cryptocurrency offers a genuine distributed purchase allowing the customer control.

\subsection{Stablecoins}

The rise of a number of stable coins now allows the ability to cash in and out of currencies when the market fluctuates. This solves a problem that would see ill-gotten gains fluctuate if remaining in the unstable currency. Previously the funds would be required to be cashed out quickly to protect the value of the asset. Stable coins allow the user to bank the value, the risk then lies with the validity of the asset \cite{2018Crypterium}. If the stable coin suffers a crash then the value is clearly lost. 

\subsection{Privacy}

A number of coins are using complex SNARK (zero-knowledge succinct non-interactive argument of knowledge) and STARK technologies to complete cryptographically strong transactions. These include value transfers and more complex smart contract transactions. Zcash and Monero are renowned for their privacy enhancing algorithms and novel ring signatures. Zcash, however, has seen a decline in the number of private transactions operated, with most of the transactions remaining transparent \cite{Rossow2018Stability}. In contrast, Monero has been lauded as the criminals go to choice for the privacy-centric coin \cite{Fanusie2018Unchained:}. There are a number of reports and instances that reference Monero but none more impactive than the WannaCry ransomware attack that hit the NHS causing countrywide disruption \cite{ESTEVES2018Monero}, \cite{ESTEVES2018Monero}.

Monero hides the sender using a ring signature, and the receiver using a stealth address. A ring signature is a digital signature that is created by a member of a group which each have their own keys. It is then not possible to determine the person in the group who has created the signature. The method of ring signatures was initially created by Ron Rivest, et al 2001 \cite{rivest2001leak}, and in their paper, they proposed the White house leak dilemma. To hide the recipient, Bob - who is part of the ring - initiates a conversation with Victor, after which Victor will know the address which Bob will use to send the transaction to. This conversational creates a new private key for Victor and an associated public address. Bob will then send his transaction to Victor to this newly created public address. Victor will then have the new private key which can then be used to transfer the funds to another account.

\subsubsection{Creating the ring}

In a ring signature, we define a group of entities who each have their own public/private key pairs of (P1, S1), (P2, S2),...,(Pn, Sn). If we want an entity i to sign a message (message), they use their own secret key (si), but the public keys of the others in the group (m,si,P1...Pn). It should then be possible to check the validity of the group by knowing the public key of the group, but not possible to determine a valid signature if there is no knowledge of the private keys within the group.

\begin{figure}
  \includegraphics[width=\linewidth]{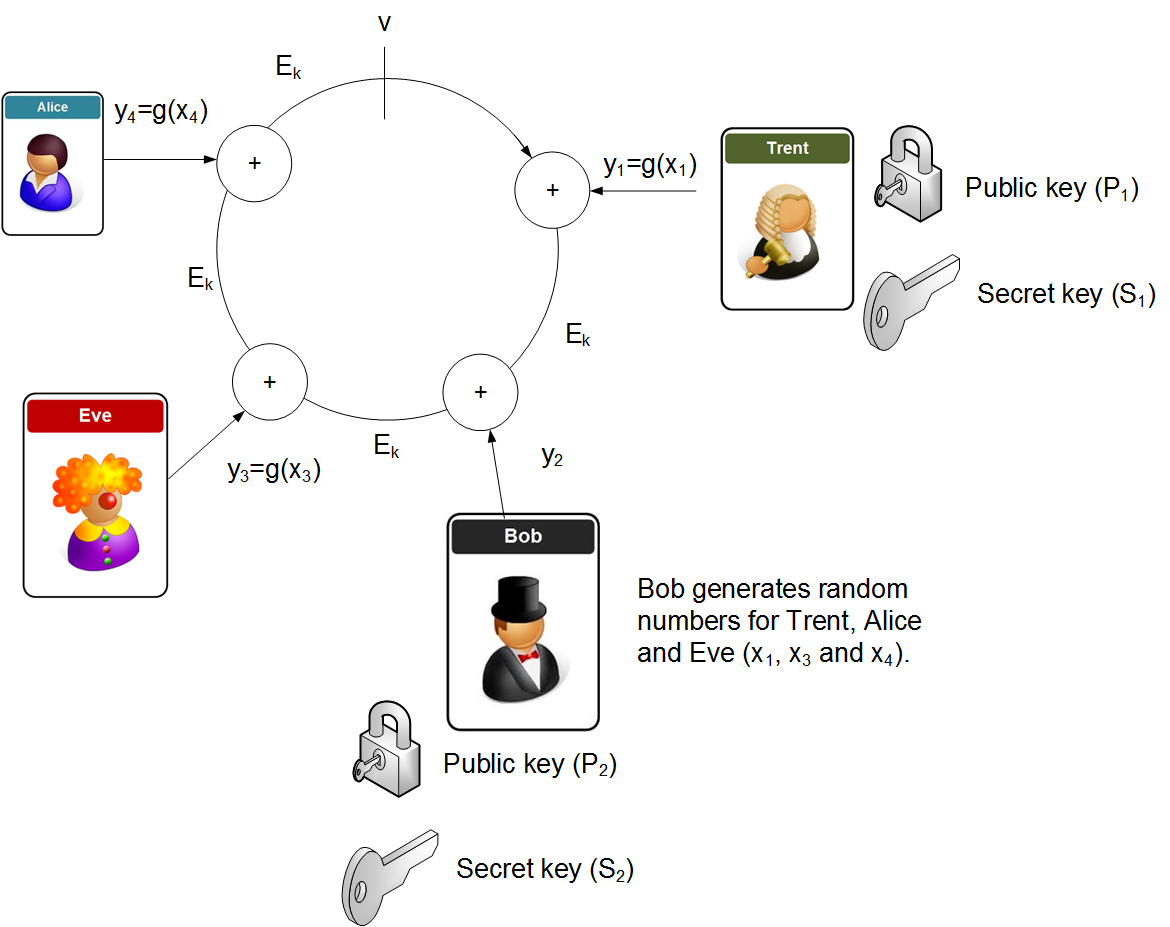}
  \caption{Ring signature}
  \label{fig03}
\end{figure}

So let’s say that Trent, Bob, Eve and Alice are in a group, and they each have their own public and secret keys. Bob now wants to sign a message from the group. He initially generates a random value v, and then generates random values (xi) for each of the other participants, but takes his own secret key (si) and uses it to determine a different secret key, and which reverse of the encryption function. He now takes the message and takes a hash of it, and thus creates a key (k). This key will be used with symmetric encryption to encrypt each of the elements of the ring (Ek), and then each element of the ring uses an EX-OR function from the previous element (Figure \ref{fig03}).

Each of the random values for the other participants is then encrypted with the public key of the given participant. Bob then computes the value of ys in order to create the ring (the result of the ring must equal v). He will then inverse this value to produce the equivalent private key (xs). Bob now releases the overall signature, and the random x values, along with the computed secret key. To check the signature, the receive just computes the ring and checks that the result matches the sent signature. 

\subsection{RingCT}

Monero initially adopted the Borromean ring signature \cite{maxwell2015borromean}, but have since migrated to a new method: Multi-layered Linkable Spontaneous Anonymous Group signature. This method hides the transaction amount and the identity of the payer and recipient. It is now known as RingCT (Ring Confidential Transactions) and was rolled-out in January 2017 and mandatory for all transactions from September 2017 \cite{noether2015ring}. For law enforcement, the implementation of RingCT makes the usage of cryptocurrency transactions increasingly difficult, as they hide both the sender and recipient of the transaction.

\section{The Solution?}

Our current financial infrastructure has existed for centuries and integrates with global and national financial regulations. A key focus for these regulations is often around anti-money laundering (AML), gathering taxes, and in the detection of financial fraud. Financial organisations must thus report on suspicious transactions, or where there are investigations on customers. Within a cryptocurrency world, there can often be little traces of financial transactions, and this is a major concern of many governments and law enforcement agencies around the world. Some criminals have even defended their actions against criminal activity by defining that the transactions were not of a financial nature \cite{mcleod2017bitcoin}.

\begin{figure}
  \includegraphics[width=\linewidth]{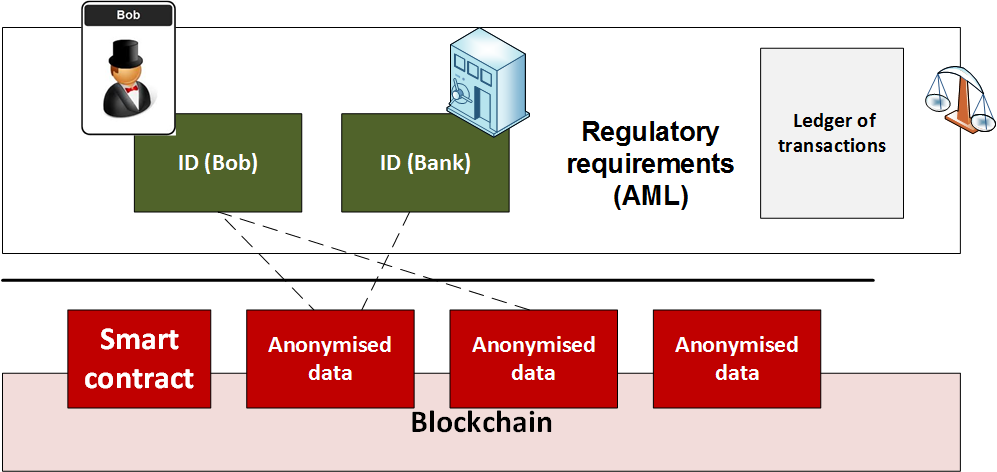}
  \caption{Regulatory mapping}
  \label{fig04}
\end{figure}

In the blockchain ecosystem, there are significant efforts to anonymise cryptocurrency transactions, which include cryptocurrencies such as zCash and Monero. This anonymisation is required with the rise of public blockchains and the increasing regulatory requirements for privacy and consent on the blockchain. Current methods of anonymisation include the usage of ring signatures and stealth addresses. The solution to this problem is to increasingly focus on anonymising the blockchain layer, but to map regulatory and statutory environments on top of the anonymisation layer, and which will provide audit trails with the revealing of the mapping from the regulatory layer into the anonymisation layer. In Figure \ref{fig04} we see an anonymised layer within the blockchain infrastructure and where the transaction sources and destinations are hidden, but the upper-level layer is then defined where real identifies will then be mapped into the anonymised infrastructure. The investigation would thus happen at the upper layer, and where the regulatory infrastructure in a country would define that all transactions would be logged from an anonymised identity to a real identity.

\section{Conclusions}
The strive for anonymisation within the blockchain infrastructure will continue, and thus the long term goal must be to start to regulate for the mapping of real identity into anonymised ones. The software which produces the transactions will thus have to keep a track for the mapping of a sovereign identity to anonymised one. Only with strong cryptography can we make sure that this is implemented in a trusted way.

\bibliographystyle{IEEEtran}
\bibliography{main1}

\end{document}